\documentclass[fleqn,usenatbib]{mnras}

\usepackage{newtxtext,newtxmath}

\usepackage[T1]{fontenc}
\usepackage{ae,aecompl}

\usepackage{graphicx}	
\usepackage{amsmath}	
\newcommand{\angstrom}{\text{\normalfont\AA}}

\defcitealias{Liu2019}{LH19}

\title[Dust attenuation at high redshift]{Geometry effects on dust attenuation curves with different grain sources at high redshift}

\author[Y.-H. Lin et al.]{
Yen-Hsing Lin,$^{1,2}$\thanks{E-mail: julius52700@gapp.nthu.edu.tw}
Hiroyuki Hirashita,$^{2}$
Peter Camps$^{3}$
and Maarten Baes$^{3}$
\\
$^{1}$Interdisciplinary Program of Science (IPS), National Tsing Hua University, No.\ 101, Section 2, Kuang-Fu Road, Hsinchu, 30013, Taiwan\\
$^{2}$Institute of Astronomy and Astrophysics, Academia Sinica,
Astronomy-Mathematics Building, No.\ 1, Section 4,
Roosevelt Road, Taipei 10617, Taiwan\\
$^{3}$Sterrenkundig Observatorium, Universiteit Gent, Krijgslaan 281 S9, B-9000 Gent, Belgium
}

\date{Accepted XXX. Received YYY; in original form ZZZ}

\pubyear{2021}

\begin{document}
\label{firstpage}
\pagerange{\pageref{firstpage}--\pageref{lastpage}}
\maketitle

\begin{abstract}
Dust has been detected in high-redshift ($z>5$) galaxies but its origin is still being debated. Dust production in high-redshift galaxies could be dominated by stellar production or by accretion (dust growth) in the interstellar medium. Previous studies have shown that these two dust sources predict different grain size distributions, which lead to significantly different extinction curves. 
In this paper, we investigate how the difference in the extinction curves affects the dust attenuation properties of galaxies by performing radiative transfer calculations.
To examine the major effects of the dust--stars distribution geometry, we adopt two representative cases in spherical symmetry: the well-mixed geometry (stars and dust are homogeneously mixed) and the two-layer geometry (young stars are more concentrated in the centre). 
In both cases, we confirm that the attenuation curve can be drastically steepened by scattering and by different optical depths between young and old stellar populations, and can be flattened by the existence of unobscured stellar populations. We can reproduce similar attenuation curves even with very different extinction curves. Thus, we conclude that it is difficult to distinguish the dust sources only with attenuation curves. However, if we include information on dust emission and plot the IRX (infrared excess)--$\beta$ (ultraviolet spectral slope) relation,
different dust sources predict different positions in the IRX--$\beta$ diagram. A larger $\beta$ is preferred under a similar IRX if dust growth is the dominant dust source.
\end{abstract}

\begin{keywords}
dust, extinction -- galaxies: evolution -- galaxies: high-redshift -- galaxies: ISM -- methods: numerical -- radiative transfer

\end{keywords}




\section{Introduction}

Dust plays an important role in various aspects of galaxy formation and evolution. Dust surfaces are the main formation site of some molecules, especially H$_2$ \citep[e.g.,][]{Gould1963,Cazaux&Tielens2004,Chen&Hirashita2018}, and are important in shielding ultraviolet (UV) heating radiation. These effects of dust surfaces keep the physical condition of dense clouds favourable for star formation.
 Dust grains also receive radiation pressure from stars and change the structure of H \textsc{ii} regions \citep[e.g.,][]{Akimkin2015,Ishiki2018} and galactic discs \citep[e.g.,][]{Ferrara1993} through dust--gas coupling. Strong radiation pressure can cause galactic outflows \citep[e.g.,][]{Murray2011}.
When dust grains absorb UV radiation, they emit photo-electrons and contribute to the heating of the surrounding gas \citep[e.g.,][]{Tielens2005}.
Dust also affects the thermal evolution of gas clouds through dust cooling.
The characteristic mass of the final fragments in star-forming clouds is regulated by dust cooling
\citep{Omukai2005}; through this process, dust cooling affects the stellar initial mass function (IMF) \citep{Schneider2006}.
 Furthermore, dust shapes the spectral energy distribution (SED) of galaxies by attenuating stellar light and reprocessing it in the far-infrared (FIR) \citep[e.g.,][]{Silva1998, Takagi1999}.
Since SEDs are used to infer physical properties of galaxies such as star formation rate, stellar mass, etc., correctly understanding the effect of dust is crucial in the study of galaxy evolution. All the above effects of dust indicate that clarifying the origin and evolution of dust is of fundamental importance in the understanding of galaxy evolution.

In recent years, dust has been detected in high-redshift ($z \gtrsim 5$) galaxies using the Atacama Large Millimetre/submillimetre Array (ALMA) \citep[e.g.,][]{Capak2015,Watson2015}. By observing dust continuum together with rest-frame FIR emission lines (e.g. [C\,\textsc{ii}] 158~$\micron$ and [O\,\textsc{iii}] 88~$\micron$), ALMA has provided us with powerful tracers to understand the physical properties of high-redshift galaxies \citep[e.g.][]{Capak2015,Inoue2016}.
The current frontier of dust observation has been extended to $z>7$ by ALMA \citep{Watson2015,Laporte2017,Hashimoto2019,Tamura2019}.
Nonetheless, the origin of dust in high-redshift galaxies is still being debated. To understand the dust enrichment at high redshift, it is crucial to disentangle the dominant source of interstellar dust. The first dust in a galaxy is supplied by stellar sources, especially supernovae (SNe) \citep{Todini2001,Nozawa2003}. The dust mass also increases through the accretion of gas-phase metals in the dense ISM \citep{Dwek1998,Draine:2009ab}. \citet{Mancini2015} showed the importance of dust growth by accretion in explaining the dust mass of a Lyman break galaxy (LBG) at $z=7.5$. \citet{Wang2017} suggested in addition that a high dust condensation efficiency in stellar ejecta gives another solution to the high dust mass in ALMA-detected galaxies at $z>6$ if SN shocks do not destroy much dust \citep[see also][]{Lesniewska2019,Nanni2020}.

To identify the dominant source of dust, we need to clarify which of the above two major sources of dust (stellar dust production and dust growth by accretion) is more efficient in observed high-redshift galaxies.
\citet[][hereafter \citetalias{Liu2019}]{Liu2019}, using the evolution model of grain size distribution developed by \citet{Hirashita2019}, showed that the resulting grain size distribution depends strongly on the dominant dust source in the following way: If the dust abundance is dominated by stellar dust production, the dust grains are predominantly large (sub-micron). If the increase of dust mass is predominantly caused by accretion, in contrast, small ($\sim 0.01~\micron$) grains dominate the dust abundance. As a consequence, the steepness of the extinction curve (the wavelength dependence of the dust extinction) is very different between these two cases. Thus, \citetalias{Liu2019} suggested that, if we can constrain extinction curves in high-redshift galaxies, we are in principle able to distinguish the dominant dust sources \citep[see also][]{Yajima2014}.

However, measurements of the extinction curve are only possible when a bright enough light source is located behind the medium of interest, and information on the intrinsic spectrum of that background source at rest-frame UV--optical wavelengths is well known. This is most ideally realized when a bright point source, such as a quasar or a gamma-ray burst, is used as background light \citep[e.g.][]{Maiolino2004,Gallerani2010,Zafar2011}. Nonetheless, the chance of finding such a bright UV--optical background source is small at $z\gtrsim 7$.
Alternatively, we can measure attenuation curves, which give the wavelength dependence of the net extinction effect of the whole galaxy \citep{Calzetti2001}. The net extinction effect -- attenuation -- is quantified by comparing the intrinsic stellar SED with the observed SED; thus, it includes all the radiative transfer effects within the galaxy \citep[e.g.][]{Witt1996,Witt2000,Granato2000}. In general, the geometry of the spatial dust--stars distribution, in addition to the dust properties, strongly affects the attenuation curve (wavelength dependence of dust attenuation) \citep[e.g.][]{Narayanan2018}. This means that, if we are able to obtain only an attenuation curve, not an \textit{original} extinction curve, it is not obvious whether we can distinguish the dominant dust sources (or dust properties) as argued above. 

The following radiative transfer effects are important in shaping the attenuation curve \citep[see][for a recent review]{Salim2020}:
(i) \textit{Geometry effect} -- UV photons from massive young stars can escape more easily if the dust-distribution geometry is complex (e.g.\ not a simple screen). This effect flattens the attenuation curve \citep[][]{Narayanan2018}. 
(ii) \textit{Age effect} -- Young stars tend to be embedded in dense regions, which enhances UV extinction \citep{Charlot2000}. This steepens the attenuation curve \citep{Narayanan2018}. 
(iii) \textit{Scattering effect} --  Scattering can affect attenuation in two ways. 
At short wavelengths, the scattering cross section of dust is large; therefore photons can experience multiple scattering before escaping from the galaxy. This effectively increases the path length of short-wavelength light within the dusty medium. As a consequence, the effective optical depth, which is proportional to the path length, is enhanced at short wavelengths \citep{Goobar2008}. 
At long wavelengths, scattering can bring photons into the line of sight, effectively decreasing the attenuation at long wavelengths. Both of these scattering effects
steepen the attenuation curve \citep{Witt2000,Baes2001}

It is not yet easy to directly measure attenuation curves, especially for high-redshift galaxies, because we need knowledge of intrinsic stellar SEDs. There are two main methods to derive attenuation properties of high-redshift galaxies \citep[see][for statistical modelling]{Mancini2016}. One of them is SED fitting \cite[e.g.][]{Papovich2001,Buat2011,Conroy2013,Kriek2013,Leja2017,Salim2018}, and the other is the IRX--$\beta$ relation \citep[e.g.][]{Meurer1999,Johnson2007,Casey2014}. IRX (infrared excess) is the FIR-to-UV flux ratio and $\beta$ is the UV spectral slope. Different attenuation curves predict different IRX--$\beta$ relations \citep[e.g. ][]{Burgarella2005, Buat2012, Salim2019}. Local starburst galaxies follow a clear sequence on the IRX--$\beta$ diagram \citep[e.g.][]{Meurer1999,Buat2005,Takeuchi2012}.
However, recent studies show that high-redshift galaxies have heavily scattered IRX--$\beta$ relations \citep[e.g.][]{Capak2015,Faisst2017,Hashimoto2019,Fudamoto2020}.
As shown by \citet{Popping2017} and \citet{Narayanan2018IRXB}, the above radiative transfer effects together with the shape of the original extinction curve affect the IRX--$\beta$ relation.

The goal of this work is to investigate whether attenuation curves and IRX--$\beta$ relations are still differentiated by the major dust sources.
We apply simple yet representative dust--stars geometries that essentially include the above three effects (i)--(iii) to calculate the attenuation curves. The current study also provides a basis on which we interpret attenuation curves and IRX--$\beta$ relations for unknown original extinction curves (or dust sources) in high-redshift galaxies. In Section \ref{sec:Methods}, we explain the models for the dust--stars geometry and the methods for the radiative transfer calculations. In Section \ref{sec:Result}, we show the resulting attenuation curves and IRX--$\beta$ relations. In Section \ref{sec:discussion}, we further discuss physical interpretations of the results. In Section \ref{sec:Conclusino}, we summarize our findings.


\section{Methods}\label{sec:Methods}

We calculate attenuation curves and IRX--$\beta$ relations based on the grain size distributions predicted for high-redshift ($z>7$) galaxies by \citetalias{Liu2019}.
First, we build a galaxy model that sets the distribution of stars and dust (dust--stars geometry). Secondly, we perform radiative transfer calculations using \textsc{skirt9}\footnote{\url{http://www.skirt.ugent.be/skirt9/}}
\citep{Camps&Baes2020A&C} and extract the attenuation curve by comparing the intrinsic stellar SED and the emergent galaxy SED. We further output the IRX--$\beta$ relations.

\subsection{Dust model}\label{subsec:DustModel}

We adopt the dust evolution model from \citetalias{Liu2019} but use the \citet{Chabier2003} IMF instead of the Salpeter IMF. We treat the galaxy as a one-zone object, and calculate the evolution of the grain size distribution by taking into account metal enrichment and interstellar dust processing \citep{Hirashita2019}. The star formation time-scale ($3\times 10^8$~yr) and the baryonic mass ($10^{10}$~M$_{\sun}$) are chosen to broadly reproduce the observed stellar masses and dust masses of galaxies at $z>7$ detected by ALMA.

\begin{figure}
	\centering
	\includegraphics[width=\columnwidth]{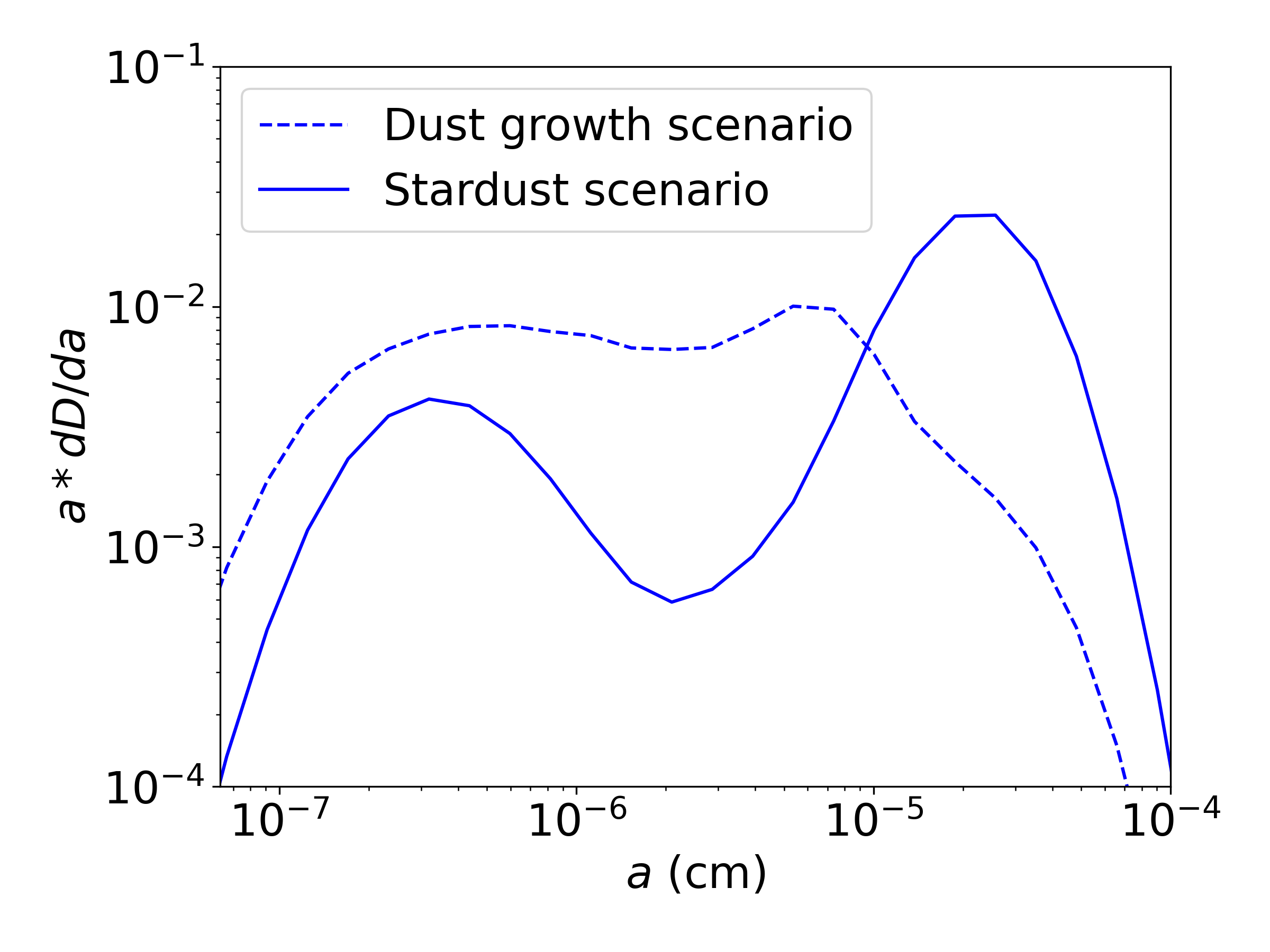}
	\includegraphics[width=\columnwidth]{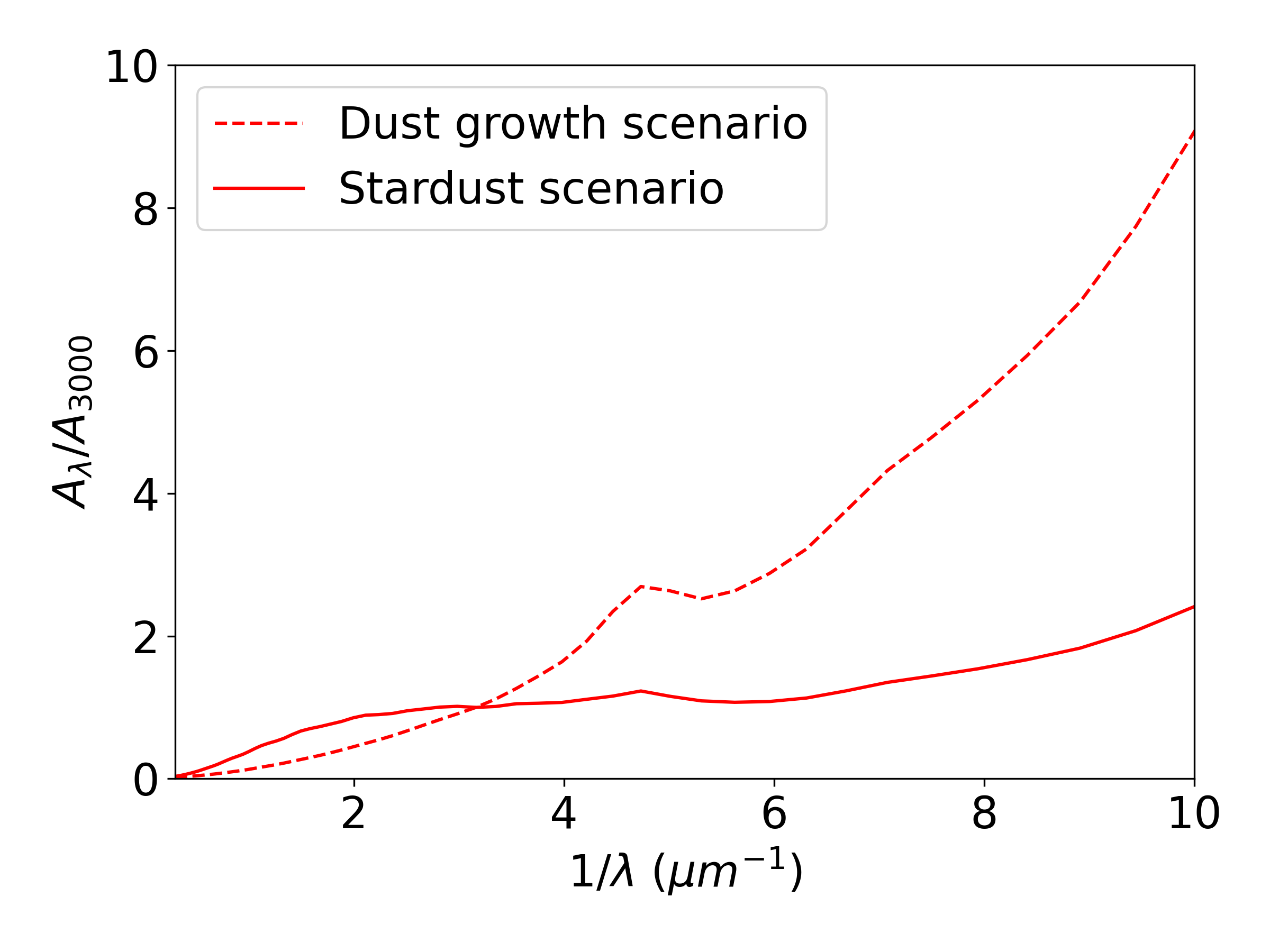}
	\includegraphics[width=\columnwidth]{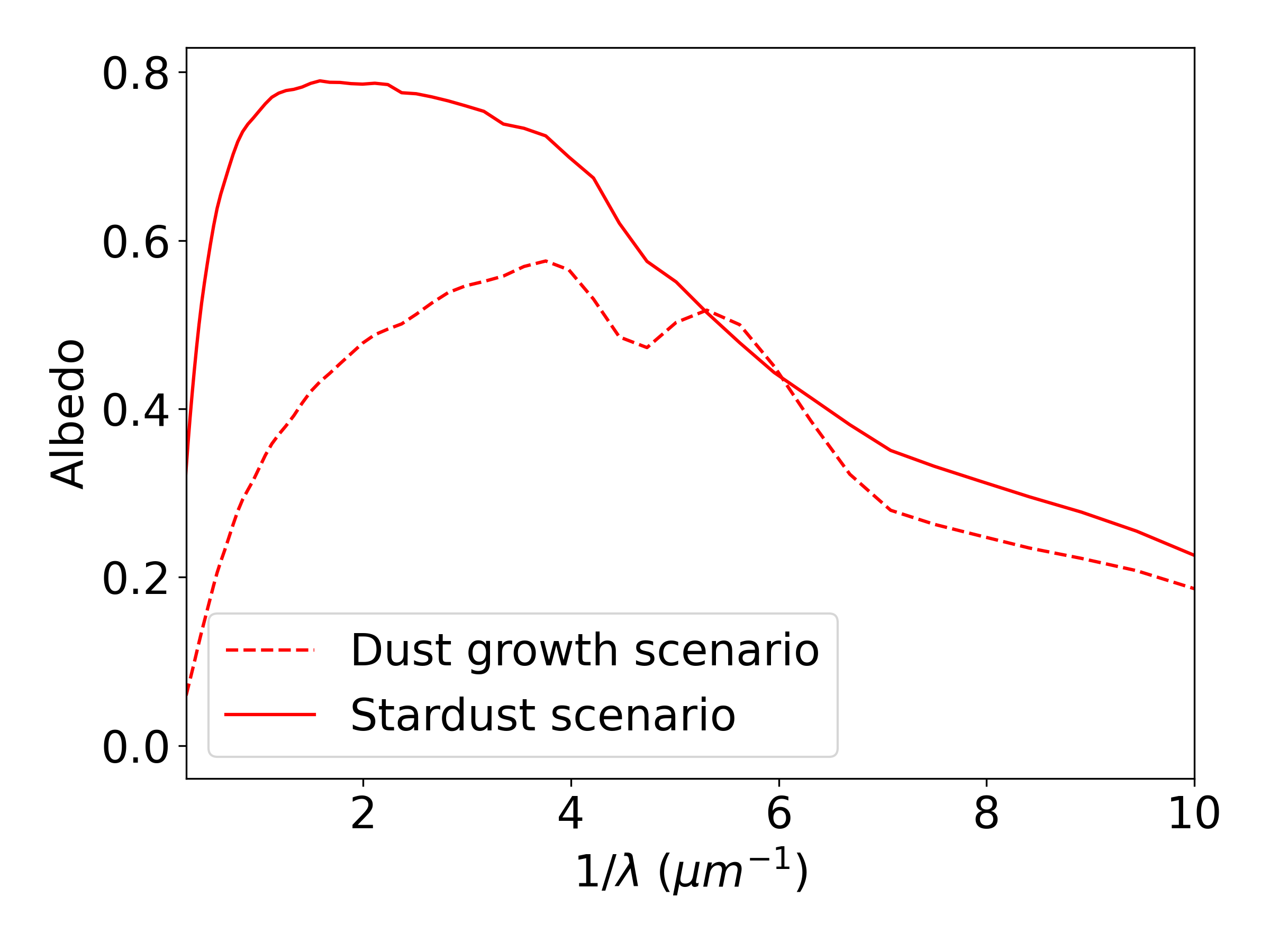}
    \caption{Upper: Grain size distribution represented by the dust-to-gas ratio per grain radius multiplied by the grain radius. This quantity (vertical axis) indicates the dust abundance per $\log a$ (horizontal axis).
    The absolute value of the vertical axis is not important because it is renormalized later using the total dust mass.
    The solid and dashed lines represent the stardust scenario and the dust growth scenario, respectively.
    Middle: Extinction curves for the stardust scenario and the dust growth scenario (solid and dashed lines, respectively). The extinction is normalized to the value at 3000 \AA.
    Lower: Albedo for the stardust scenario and the dust growth scenario (solid and dashed lines, respectively).}
    \label{fig:GrainSizeDis}
\end{figure}

As shown by \citetalias{Liu2019}, the resulting extinction curves are sensitive to the dominant dust production mechanisms: stellar dust production vs.\ dust growth by accretion.
These processes are regulated by the dust condensation efficiency in stellar ejecta ($f_\mathrm{in}$) and the time-scale of accretion at solar metallicity ($\tau_\mathrm{0,acc}$).
We adopt the following two sets of parameter values: $(f_\mathrm{in},\,\tau_\mathrm{0,acc}/\mathrm{yr})=(0.5,\, 1.61 \times 10^8)$, and (0.1, $1.61 \times 10^7$). The former explains the dust content of the $z>7$ galaxies mainly by stellar dust production, and the latter by accretion. These first and second scenarios are referred to as the \textit{stardust scenario} and the \textit{dust growth scenario}, respectively.
The age of the system is set to be $3\times10^8$ yr, which is roughly consistent with the ages constrained from stellar spectra for the galaxy sample (\citetalias{Liu2019}; originally from \citealt{Tamura2019, Watson2015, Laporte2017, Hashimoto2019}). The grain size distributions (the dust mass abundances per $\log a$) are shown in Fig.\ \ref{fig:GrainSizeDis} (top panel). The absolute values of these distributions are not important, because the distributions will be renormalized to match the total dust mass.

We also show the calculated extinction curves and albedo for the two scenarios in Fig.\ \ref{fig:GrainSizeDis} (middle and bottom panels, respectively).
In calculating the extinction curves, we assume a silicate--graphite mixture and use the optical dust properties taken from \citet{Draine1984} and \citet{Laor1993}.
For the dust species, based on the results in \citet{Hirashita2020}, the mass fraction of silicate at $t\sim 3\times 10^8$ yr is roughly 0.9; thus, in this paper, the mass ratio between silicate and graphite is assumed to be 9:1. As shown in Fig.\ \ref{fig:GrainSizeDis},
the extinction curve in the stardust scenario is flat because the grain size distribution is dominated by large grains, while that in the dust growth scenario is steep. The difference in the extinction curves between the two scenarios is robustly large regardless of the assumed silicate fraction. Thus, the discussions and conclusions in this paper are not sensitive to the adopted silicate fraction. 
In both scenarios, the albedo increases towards longer wavelengths, which enhances scattering at longer wavelengths. We discuss the effect of the albedo on the slope of the attenuation curve in Section \ref{subsec:steepening}.

\subsection{Galaxy model}\label{subsec:galaxy}

\citet{Narayanan2018} discussed two main effects that change the slope of the attenuation curve under a given extinction curve, as described in the Introduction. The first effect is the geometry effect caused by nonuniform extinction for stars (light from some stars escapes easily by a geometrical effect). The second is the age effect arising from different optical depths between young and old stellar populations (young stars are usually more embedded in dense regions) as investigated by \citet{Silva1998}, \citet{Bianchi2000}, and \citet{Inoue2005}.
They showed that the geometry effect makes the attenuation curve flatter, while the age effect makes it steeper. These two effects are regulated by the spatial distribution of stars relative to that of dust.

Since little is known about the actual geometry in high-redshift galaxies, we adopt a geometry model that is simple (here, spherically symmetric) but still capable of examining the above two effects (geometry and age effects). As explained later, these two effects can, in practice, be examined using spherically symmetric distributions. The simple assumption also makes the computations easier and the parameter search physically transparent. Because spherical geometry already contains basic radiative transfer effects, it has been adopted in previous studies \citep[e.g.,][]{Witt1992,Witt1996,Witt2000}.
We leave the effect of realistic non-spherical or complex structures, which could be treated by post-processing simulated galaxies \citep[e.g.,][]{Trcka2020, Vogelsberger2020} for future work.

\begin{figure}
    \centering
    \includegraphics[width=0.45\textwidth]{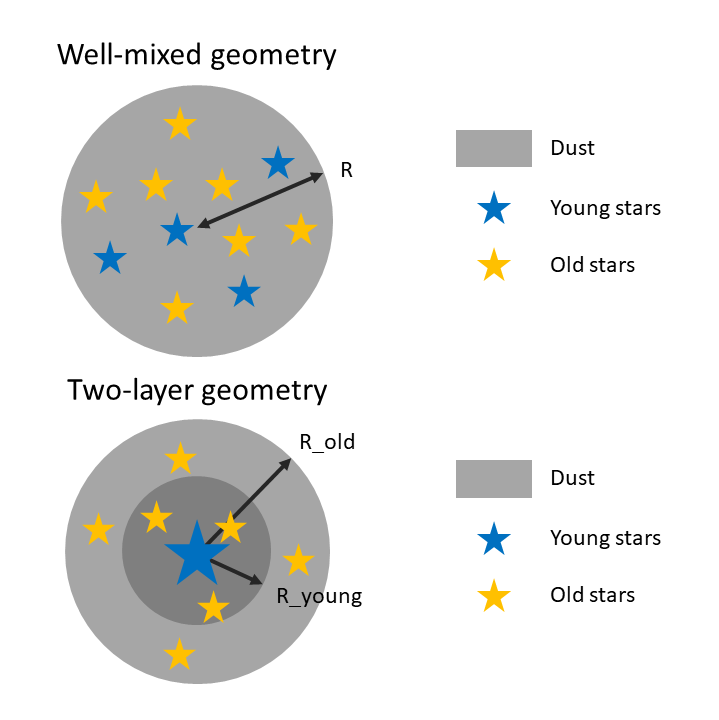}
    \caption{
    Two types of geometry: well-mixed geometry and two-layer geometry. Both are spherically symmetric. In the first case, all stars and dust are distributed uniformly within a sphere of radius $R$ while in the second case, young stars are concentrated in the centre surrounded by a denser region with radius $R_\mathrm{young}$, and old stars are distributed uniformly within $R_\mathrm{old}$.}
    \label{fig:Model}
\end{figure}

First, we set up a model of a sphere in which stars and dust are distributed homogeneously within the same radius (i.e.\ stars and dust are well mixed). This is referred to as the \textit{well-mixed geometry} (Fig.~\ref{fig:Model}). In this geometry, some stars are always located near the outer boundary, and their radiation easily escapes. Thus, in this well-mixed geometry, we realize the above geometry effect.
Secondly, we adopt another spherically symmetric model in which we separate young and old stellar populations by concentrating the young population in the centre. This \textit{two-layer geometry} (Fig.~\ref{fig:Model}) realizes the age effect; that is, young stars suffer more extinction than old stars. As shown later, these two geometries actually produce consistent behaviours of attenuation curves with those obtained by \citet{Narayanan2018}.
The detailed setup of the above two geometry models is explained in what follows.

\subsubsection{Well-mixed geometry}
In the well-mixed geometry, we simply adopt a uniform sphere with radius $R$ with uniformly distributed stars and dust of mass $M_*$ and $M_\mathrm{d}$, respectively. The stellar mix is composed of young and old stellar populations with ages $3\times 10^6$ and $3\times 10^8$ yr, respectively. The age of the old stellar population is the same as the galaxy age adopted above, and the young stellar population is assumed to occupy 1 per cent of the total stellar mass in proportion to the assumed ratio of the age. The existence of a stellar population as young as a few Myr is indeed indicated by stellar population analysis of the $z>7$ galaxies \citep{Tamura2019,Hashimoto2019}. The detailed stellar age setup does not affect our conclusions as long as the far-UV ($\lambda\sim 0.15~\micron$) and near-UV ($\lambda\sim 0.3~\micron$) emission is dominated by stellar populations with different ages.
We discuss the effect of stellar age in Section \ref{subsec:clumps}.
Considering that the galaxies are in the early phase of evolution, we adopt a low stellar metallicity $Z\sim 1.2\times 10^{-3}$ ($\sim 0.1$ Z$_{\sun}$); however, the stellar metallicity has much less influence on the results than other parameters.

The dust mass and the galaxy radius are degenerate in deciding the optical depth of the system. Thus, we only change $R$. We fix $M_\mathrm{d}$ to $1.5\times 10^6$ M$_{\sun}$ and
$M_*$ to $1.5\times 10^9$ $\textrm{M}_{\sun}$ according to the sample adopted by \citetalias{Liu2019}. Although the diversity of stellar mass among the observation data is large, it has almost no influence on the attenuation curves and the IRX--$\beta$ relation. 

The total optical depth varies with $R$. The spatial extent of dust emission is comparable to the resolution of ALMA ($\sim$a few kpc)  \citep[e.g.][]{Fujimoto2017,Hashimoto2019,Novak2020}. We examine $R=5$ kpc for a case of moderate extinction, and shrink it down to 0.5 kpc for a case where stars are heavily obscured. 
We later confirmed that the dust temperatures in this radius range lie between 25 and 40 K. These dust temperatures were obtained by comparing the peak wavelength of each simulated FIR SED with the analytically expected peak position in a modified black body SED $\propto\nu^\beta B(\nu,\, T)$ with $\beta=2$ [$B(\nu,\, T)$ is the Planck function at frequency $\nu$ and temperature $T$]. This can be directly interpreted as an observed (or luminosity-weighted) dust temperature (`effective dust temperature' in \citealt{Liang2019}). The temperature range is roughly consistent with that assumed in the above mentioned observational studies. Nevertheless, since the dust temperatures are poorly constrained for high-redshift galaxies, we do not use the dust temperatures to constrain the model. The parameters are summarized in Table \ref{tab:Uni_Parameters}.

\begin{table}
    \caption{Parameters of the well-mixed geometry.}
	\label{tab:Uni_Parameters}
	\begin{tabular}{lll} 
		\hline
		Parameter & Meaning & Value \\
		\hline
		$R$ & Radius of the galaxy & 0.5--5 kpc  \\
		$M_{\mathrm{d}}$ & Mass of the dust in the galaxy & $1.5 \times10^{6}$ M$_{\sun}$\\
		$M_{\mathrm{*,old}}$ & Stellar mass of old stars & $1.5\times10^{9}$ M$_{\sun}$  \\
		$M_{\mathrm{*,young}}$ & Stellar mass of young stars & $1.5\times10^{7}$ M$_{\sun}$  \\
		$T_{\mathrm{old}}$ & Age of the old stars & $3\times10^8$ yr  \\
		$T_{\mathrm{young}}$ & Age of young stars & $3\times10^6$ yr  \\
		\hline
	\end{tabular}
\end{table}

\subsubsection{Two-layer geometry}
In the two-layer geometry, we aim at examining the age effect, that is, more extinction for young stars. We adopt the same ages and mass fractions for the young and old stellar populations as in the well-mixed geometry. We distribute these two stellar populations in two spheres with different radii, keeping the spherical symmetry. The two components are referred to as the young and old sphere.
The old sphere is a homogeneous sphere containing old stars and dust with masses $M_{\mathrm{d,old}}$ and $M_\mathrm{*,old}$, respectively. 
The young sphere with radius $R_{\mathrm{young}}$ contains dust with mass $M_\mathrm{{d,young}}$ and stars with mass $M_{\mathrm{*,young}}$. To maximize the age effect, we place all the young stars at the centre of the galaxy. We change $R_\mathrm{young}$ from 0.5 kpc to 5 kpc to adjust the optical depth for the young stars, while we fix $R_\mathrm{old}=5$ kpc. The adopted values of the relevant parameters are listed in Table \ref{tab:TS_Parameters}.

\begin{table}
	\caption{Parameters of the two-layer geometry.}
	\label{tab:TS_Parameters}
	\begin{tabular}{lll} 
		\hline
		Parameter & Meaning & Value \\
		\hline
		$R_{\mathrm{old}}$ & Radius of the old sphere & 5 kpc  \\
		$R_{\mathrm{young}}$ & Radius of the young sphere & 0.5--5 kpc  \\
		$M_{\mathrm{d,old}}$ & Mass of dust in the old sphere & $1.5 \times10^{6}$ M$_{\sun}$\\
		$M_{\mathrm{d,young}}$ & Mass of dust in the young sphere & $1.5\times10^{5}$ M$_{\sun}$\\
		$M_{\mathrm{*,old}}$ & Stellar mass of old stars & $1.5\times10^{9}$ M$_{\sun}$  \\
		$M_{\mathrm{*,young}}$ & Stellar mass of young stars & $1.5\times10^{7}$ M$_{\sun}$  \\
		$T_{\mathrm{old}}$ & Age of the old stars & $3\times10^8$ yr  \\
		$T_{\mathrm{young}}$ & Age of young stars & $3\times10^6$ yr  \\
		\hline
	\end{tabular}
\end{table}

\subsection{Radiative transfer calculation}\label{subsec:Radiation}

We use \textsc{skirt} \citep{Baes2003, Baes2011, Camps2015, Camps&Baes2020A&C} to calculate the radiative transfer inside the dust--stars distributions in Section \ref{subsec:galaxy}. \textsc{skirt} is a C++ based 3D Monte Carlo radiative transfer code suitable for handling dusty systems, ranging from galaxies to AGNs and circumstellar discs \citep[e.g.,][]{Stalevski2012, Saftly2015, Hendrix2016, Nersesian2020a, Nersesian2020b}. For the present work, we use the latest version of the code, \textsc{skirt9} \citep{Camps&Baes2020A&C}.

In our work, we import our galaxy model (the density fields of dust and stars). Dust and stars are represented by $N=10^4$ particles with an appropriate smoothing length. Since we change the size of the galaxy, we calculate the smoothing length $l_\mathrm{smooth}$ of the particles by the following equation:
\begin{equation}
    l_\mathrm{smooth} = R(S_0 N)^{-1/3},
\end{equation}
where $S_0$ is the smoothing parameter, which is set to 0.1. The radius $R$ and $N$ are replaced with the appropriate values in the two-layer geometry.

For the dust, we assume local thermal equilibrium
and handle the optical and calorimetric properties of graphite and silicate
using Draine's model \citep{Draine1984,Laor1993,Weingartner&Draine2000, Draine&Li2001}. To include the grain size distribution, which is particularly important for the extinction curve, we divide the grain sizes into 31 bins for each species and import them into \textsc{skirt9} as separate components. For the stars, the spectrum is synthesized using the \citet{Bruzual2003} model with the \citet{Chabier2003} IMF.

\subsection{Output quantities}\label{subsec:Output}
\subsubsection{Attenuation Curves}\label{subsubsec:attenuation}
After calculating the radiative transfer inside the galaxy, we output the intrinsic and observed SEDs. We define the effective optical depth at wavelength $\lambda$, $\tau_\lambda^\mathrm{eff}$ as 
\begin{equation}
    \tau_{\lambda}^\mathrm{eff}=\ln\left(\frac{L_{\nu , 0}}{L_{\nu , \mathrm{esc}}}\right) ,
	\label{eq:tau}
\end{equation}
where $L_{\nu ,0}$ and $L_{\nu , \mathrm{esc}}$ are the intrinsic and escaping (attenuated) luminosities per unit frequency $\nu$, respectively. The attenuation in units of magnitude, $A_\lambda$, is related to the above effective optical depth as $A_\lambda =(2.5\log_{10}\mathrm{e})\tau_\lambda^\mathrm{eff}$ \citep{Evans1994}.
The attenuation is normalized to the value at $\lambda =3000~\angstrom$ ($A_\mathrm{3000~\angstrom}$),
so that we present $A_\lambda /A_{3000~\angstrom}$ for the attenuation curve. We also compare the attenuation curves with the input (original) extinction curve shown in Fig.\ \ref{fig:GrainSizeDis} (lower).

\subsubsection{Slope--$A_{V}$--$R$ diagram}
To compare the slopes of attenuation curves for different galaxy radii quantitatively, we define the following quantity $S$ that represents the slope between rest-frame $\lambda =1500$ and 3000 \AA\ \citep{Salim2020}:
\begin{equation}
    S \equiv A_{1500~\angstrom}/A_{3000~\angstrom}.\label{eq:S}
\end{equation}
We later plot $S$ as a function of $A_V$ (attenuation in the $V$ band).

\subsubsection{IRX--$\beta$ relation}
It is still hard to directly measure attenuation curves for high-redshift galaxies. This is mainly due to the difficulty in obtaining the intrinsic stellar spectrum at a number of wavelengths. Therefore, it is useful to predict quantities that are more easily observed for high-redshift galaxies. The IRX--$\beta$ relation is often used to discuss the dust attenuation and emission properties (see the Introduction). 
The IRX--$\beta$ relation can be observationally derived if we obtain photometric measurements at two  or more rest-frame UV wavelengths in addition to an estimate of the total dust luminosity. Thus, the ALMA-detected $z>7$ LBGs discussed above have suitable data for the IRX--$\beta$ relation.
From the definition of $\beta$ ($L_\lambda\propto\lambda^\beta$), we obtain
\citep[see e.g.][]{Ouchi2013}
\begin{equation}
    \beta=
    \frac{\log_{10}(L_{\nu_2}/L_{\nu_1})}{\log_{10}({\lambda_2}/{\lambda_1})} -2,
	\label{eq:beta}
\end{equation}
where $\lambda_1$ and $\lambda_2$ are two rest-frame wavelengths at which the luminosity is measured, and $\nu_1$ and $\nu_2$ are the frequencies corresponding to these two wavelengths.
In this work, we choose $\lambda_{1}=0.16~\micron$, and $\lambda_{2}=0.25~\micron$ to avoid the effect of the 2175 \AA\ bump caused by graphite \citep[e.g.][]{Draine1984}. Since the strength of this bump in high-redshift galaxies is still unclear, we avoid theoretical predictions being affected by the bump strength. IRX is calculated as
\begin{equation}
    \mathrm{IRX}=\log_{10}\left(\frac{L_\mathrm{IR}}{L_\mathrm{UV}}\right) ,
	\label{eq:irx}
\end{equation}
where $L_\mathrm{IR}$ is the infrared luminosity integrated from rest-frame $\lambda =3~\micron$ to 1~mm, and $L_\mathrm{UV}$ is the UV luminosity estimated by $L_\mathrm{UV}=\nu_1 L_{\nu_1}$.

As references, we adopt some empirical and observational IRX--$\beta$ relations. As empirical relations we use the predictions from the Calzetti attenuation curve \citep{Calzetti2000} and the Small Magellanic Cloud (SMC) extinction curve (in this case, we assume that the attenuation curve is the same as the SMC extinction curve). For observational IRX--$\beta$ relations, we take the data for galaxies at $z \geq 6.6$ from \citet[][]{Hashimoto2019}.
These data are just used as guides or references for our results, and are not intended for detailed comparison or fitting. We also note that the observational data have large uncertainties; for example, IRX depends on the assumption on the dust temperature (50 K is adopted for the data from \citealt{Hashimoto2019}), whose 1-$\sigma$ uncertainty is typically 10 K. The value of $\beta$ depends on the choice of the observing wavelengths \citep{Popping2017, Liang2021}. Many of the LBGs at $z\gtrsim 6$ are not detected even by ALMA (e.g.\ \citealt{Bouwens2016}; also some galaxies in \citealt{Hashimoto2019}), which makes it difficult to draw firm conclusions on the IRX--$\beta$ relation at high redshift. In this paper, we focus on the effect of dust sources and dust distribution geometry from a theoretical point of view, and further comparison with observational data is left for future work. 
Nonetheless, we will see later that the IRX--$\beta$ relations we obtain are not inconsistent with the possible diversity in the currently available data from \citet{Hashimoto2019}.

We also note that the different choices of wavelengths between $S$ and $\beta$ do not affect our discussions and conclusions below as long as they are in the UV range. First of all, $S$ and $\beta$ are slope indicators of different quantities (attenuation curves and observed SEDs, respectively). Moreover, even if we adopt the wavelengths 0.15 and 0.3 $\mu$m (used for $S$) for $\beta$, the diversity in the IRX--$\beta$ diagrams is rather enhanced between the different dust scenarios; especially $\beta$ could become larger by $\sim 1$ for the dust growth scenario, while $\beta$ does not change much in the stardust scenario. Thus, adopting 0.16 and 0.25 $\mu$m gives a conservative estimate for the diversity between the two scenarios.


\section{Results} \label{sec:Result}

In this section, we show the results for the well-mixed and two-layer geometries. We examine how diverse the attenuation curves can be and which region in the IRX--$\beta$ diagram can be explained with different dust--stars geometries.

\subsection{Well-mixed geometry}\label{Res:WellMixed}

\begin{figure}
    \centering
    \includegraphics[width=0.45\textwidth]{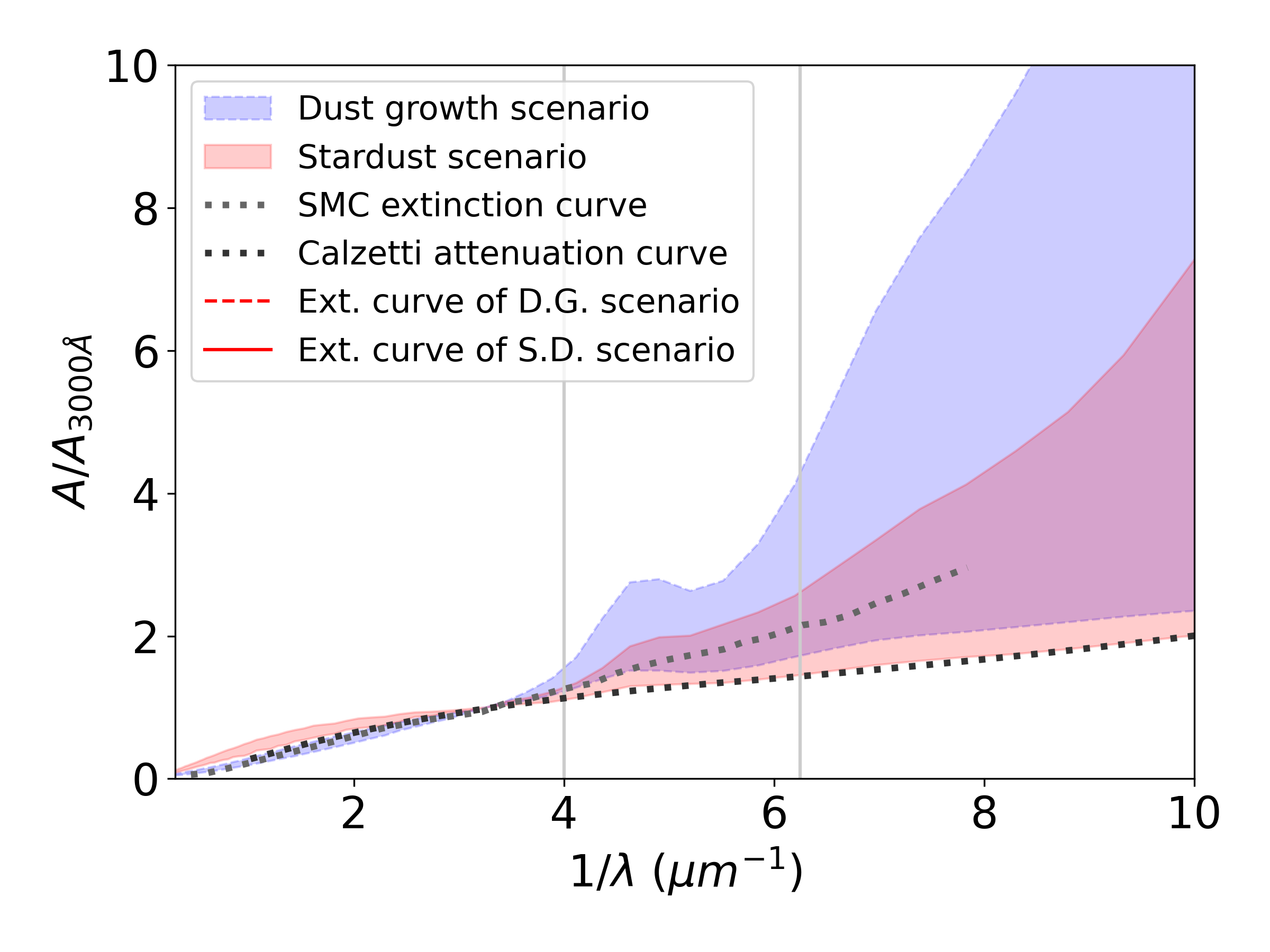}
    \includegraphics[width=0.45\textwidth]{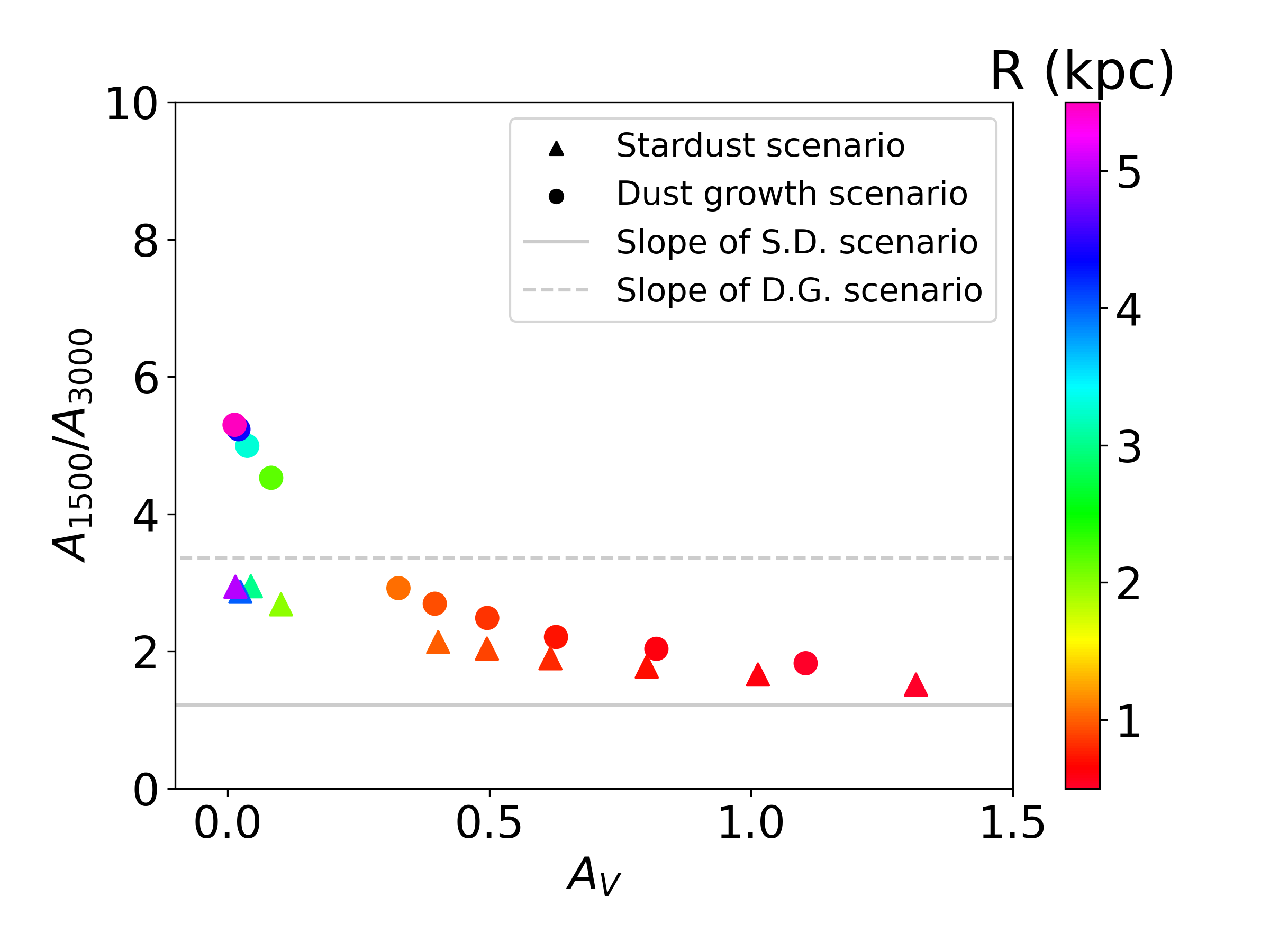}
    \includegraphics[width=0.45\textwidth]{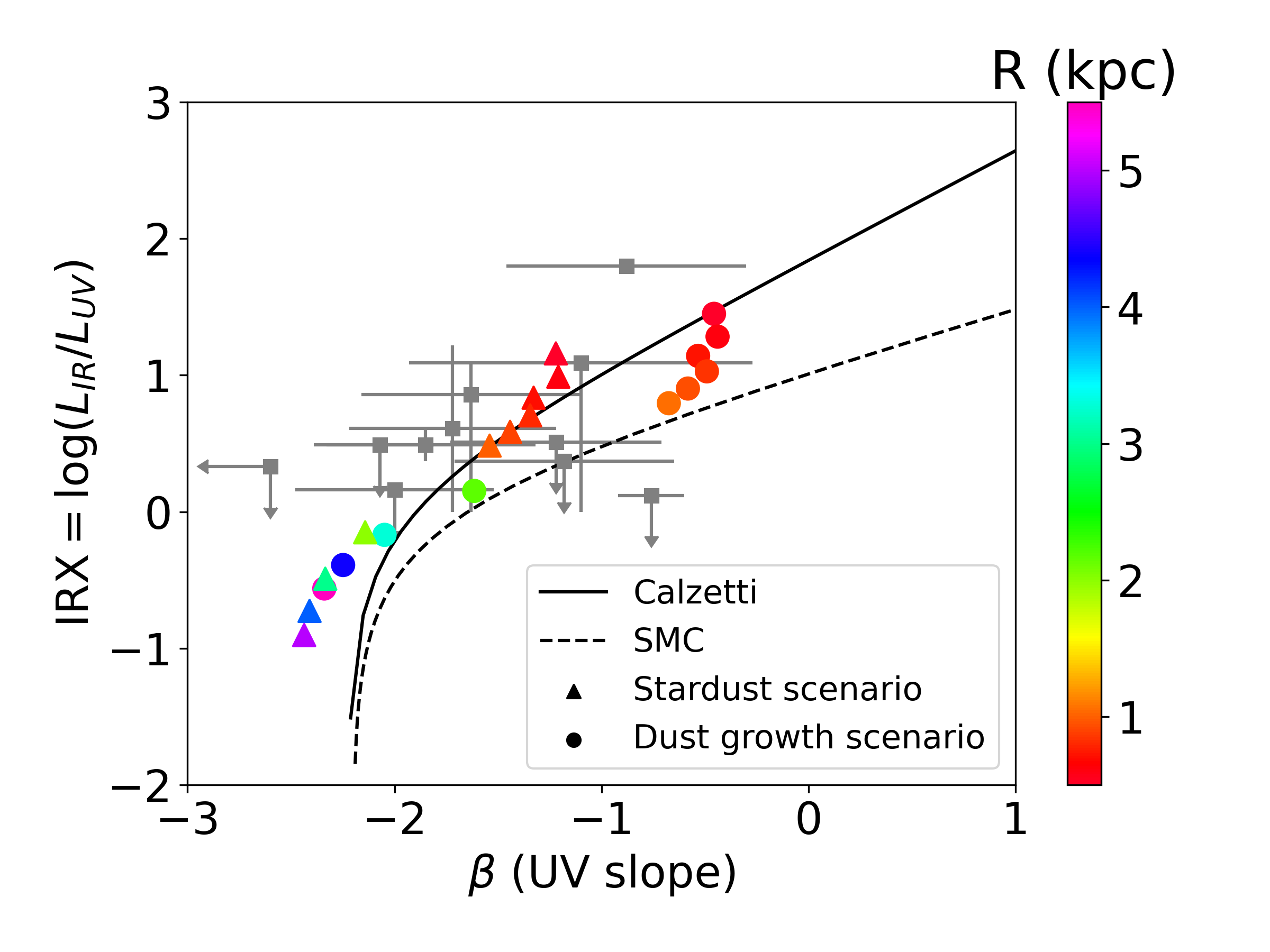}
    \caption{Results for the well-mixed geometry. 
    Top: Attenuation curves normalized to the value at $\lambda =3000$ \AA. The red region 
    (the blue region)
    represents the area covered by attenuation curves of the stardust scenario (the dust growth scenario). The upper and lower boundaries correspond to $R=5$ and 0.5~kpc, respectively. The red solid and dashed lines represent the original extinction curves of the stardust scenario and of the dust growth scenario, respectively. The vertical gray lines mark the wavelengths at which the UV slope $\beta$ is measured. We also show the Calzetti attenuation curve and the SMC extinction curve for reference by the black and gray dotted lines, respectively.
    Middle: Relation among $R$ (colour bar), $A_{V}$ and the slope of attenuation curve ($S \equiv A_{1500}/A_{3000}$). The solid and dashed lines represent the slopes of the original extinction curves in the stardust scenario and in the dust growth scenario, respectively.
    Bottom: IRX--$\beta$ diagram. The triangles and the circles represent the stardust scenario and the dust growth scenario, respectively. The colour indicates the radius of the galaxy ($R$) as shown in the colour bar. The gray squares with error bars are observational data taken from \citet[][]{Hashimoto2019} for $z\geq 6.6$ galaxies. The solid and dashed lines are the IRX--$\beta$ relations derived from the Calzetti and SMC law, respectively.
    }
    \label{fig:Uni_Att}
\end{figure}

First, we examine the well-mixed geometry.
In the top panel of Fig.\ \ref{fig:Uni_Att} we show the attenuation curves corresponding to various galaxy radii ($R$). The original extinction curves for the stardust scenario and the dust growth scenario are both presented and compared with the attenuation curves. We also show the relationship between the attenuation curve slope ($S$ in equation \ref{eq:S}) and the effective optical depth (represented by $A_V$), which is a function of $R$, in the middle panel of Fig.\ \ref{fig:Uni_Att}. 
We observe the following features:
\begin{enumerate}
    \item The attenuation curves in the dust growth scenario still inherit the steep slope from the original extinction curves, compared with those in the stardust scenario. At low optical depth (large $R$), the attenuation curves are significantly steeper than the original extinction curve. The steepness drops rapidly as the galaxy radius decreases, and the attenuation curves become less steep than the original extinction curve at $R \lesssim 1$ kpc. There are two competing mechanisms here: one is the scattering effect, which steepens the attenuation curve as mentioned in the Introduction; the other is the geometry effect (Introduction; Section \ref{subsec:galaxy}) caused by the leakage of stellar light from the outer part of the sphere, which flattens the attenuation curve. In the low optical depth regime, the former dominates and makes the attenuation curve steeper than the original extinction curve, while in the high optical depth regime, the latter effect flattens the attenuation curve. We will discuss these effects in detail in Section \ref{sec:discussion}.
    \item The attenuation curves in the stardust scenario are steepened compared with the original extinction curve. As also observed in the dust growth scenario, the slope of the attenuation curve drops as the optical depth increases. However, the slope never becomes shallower than that of the original extinction curve in the stardust scenario.
    \item Overall, the attenuation curves in the two scenarios become close and significantly overlap with each other in the range of galaxy radii examined in our models. In particular, for high optical depth (small $R$), the attenuation curves of both scenarios have similar slopes ($S$). Shallow slopes similar to the Calzetti attenuation curve are also realized. 
\end{enumerate}

In the bottom panel of Fig.\ \ref{fig:Uni_Att} we show the IRX--$\beta$ relation for the well-mixed geometry. We compare the two scenarios with different galaxy radii $R$ and observe the following features:
\begin{enumerate}
    \item In each scenario, IRX and $\beta$ increase as the galaxy radius decreases because of the increasing optical depth (attenuation). At high optical depth ($R \lesssim 2$ kpc), the points broadly follow the relations expected for the SMC or Calzetti laws. However, we find that $\beta$ stops increasing at $R\lesssim 0.6$ kpc, where the theoretical points evolve vertically on the IRX--$\beta$ diagram as $R$ decreases. The reason for this `saturation' of $\beta$ is explained in Section \ref{subsec:flattening}. A similar effect is also shown for the case of finite dust covering fraction in \citet{Popping2017}: $\beta$ cannot increase beyond a certain value if a part of the stellar UV light always escapes from the galaxy.
    \item For the same galaxy radius, the dust growth scenario has larger $\beta$ and IRX than the stardust scenario. This is due to the difference in the extinction curves. Thus, even if the attenuation curve shapes are similar, the location in the IRX--$\beta$ relation generally depends on the original extinction curve shape (see also Section \ref{subsec:flattening}). Moreover, even if the attenuation curves are only slightly different at high optical depth (Fig.\ \ref{fig:Uni_Att} upper and middle), such a slight difference causes an amplified difference in $\beta$ as explained in Section \ref{subsec:distinguish}.
\end{enumerate}

With the results above, we conclude that the well-mixed geometry has an effect of homogenizing the attenuation curves: the very flat and very steep extinction curves in the two scenarios could produce similar attenuation curves. At high optical depth, the points lie around the region predicted from the Calzetti and SMC curves because the attenuation curves become similar to them.
With the same $R$, the dust growth scenario (with a steeper extinction curve) tends to produce larger IRX and $\beta$ than the stardust scenario.

\subsection{Two-layer geometry}\label{Res:TwoLayer}

\begin{figure}
    \centering
    \includegraphics[width=0.45\textwidth]{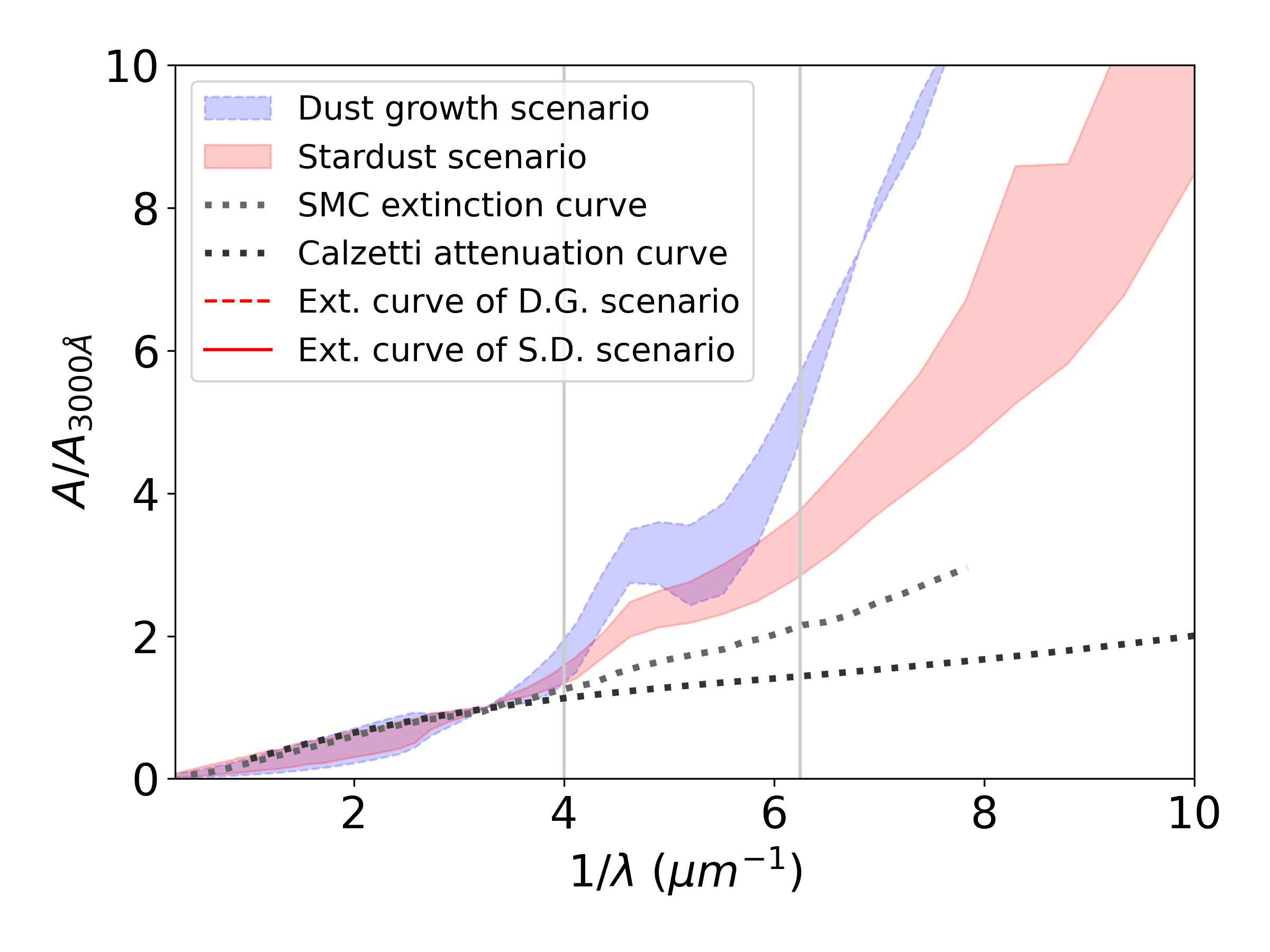}
    \includegraphics[width=0.45\textwidth]{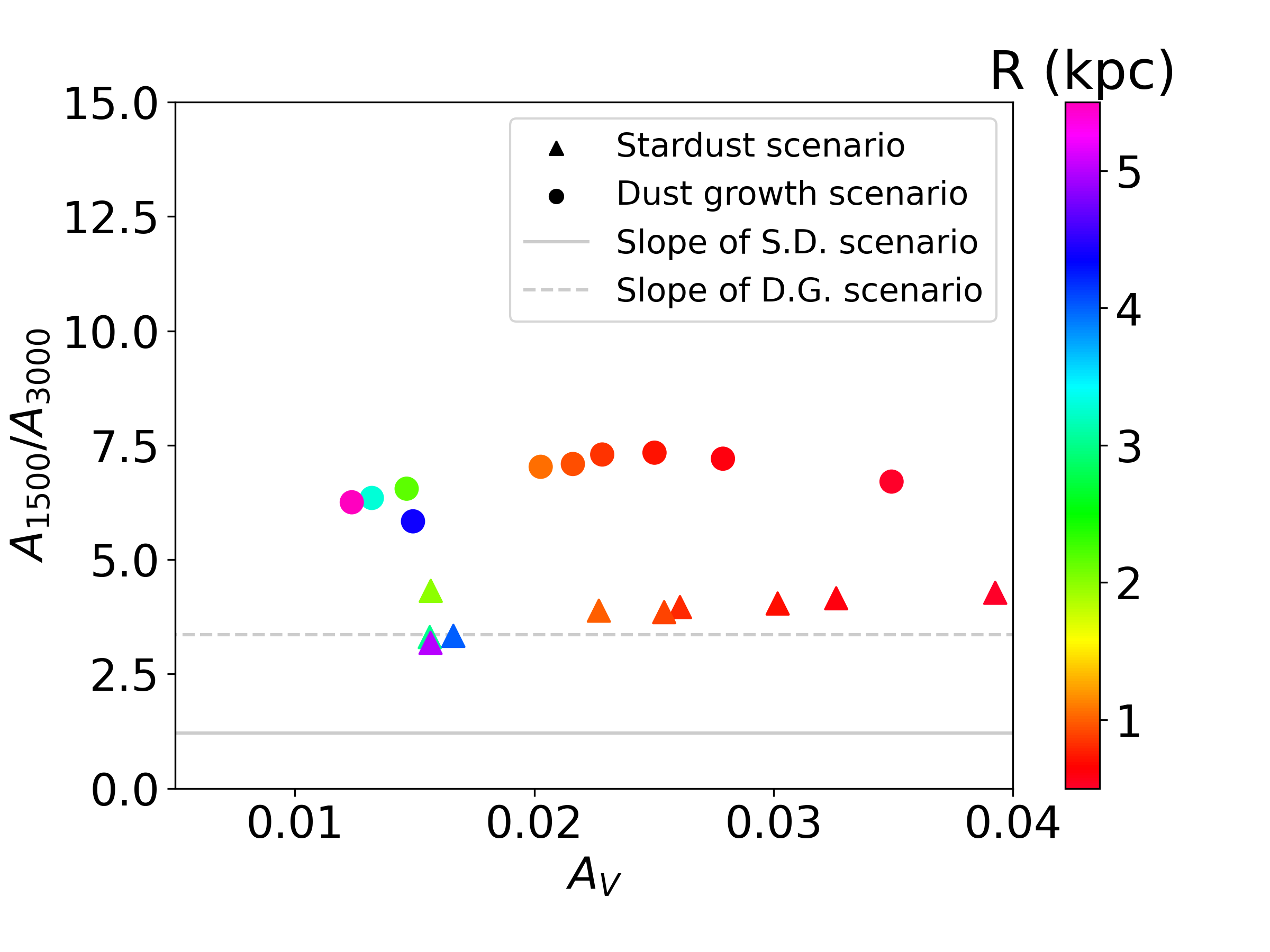}
    \includegraphics[width=0.45\textwidth]{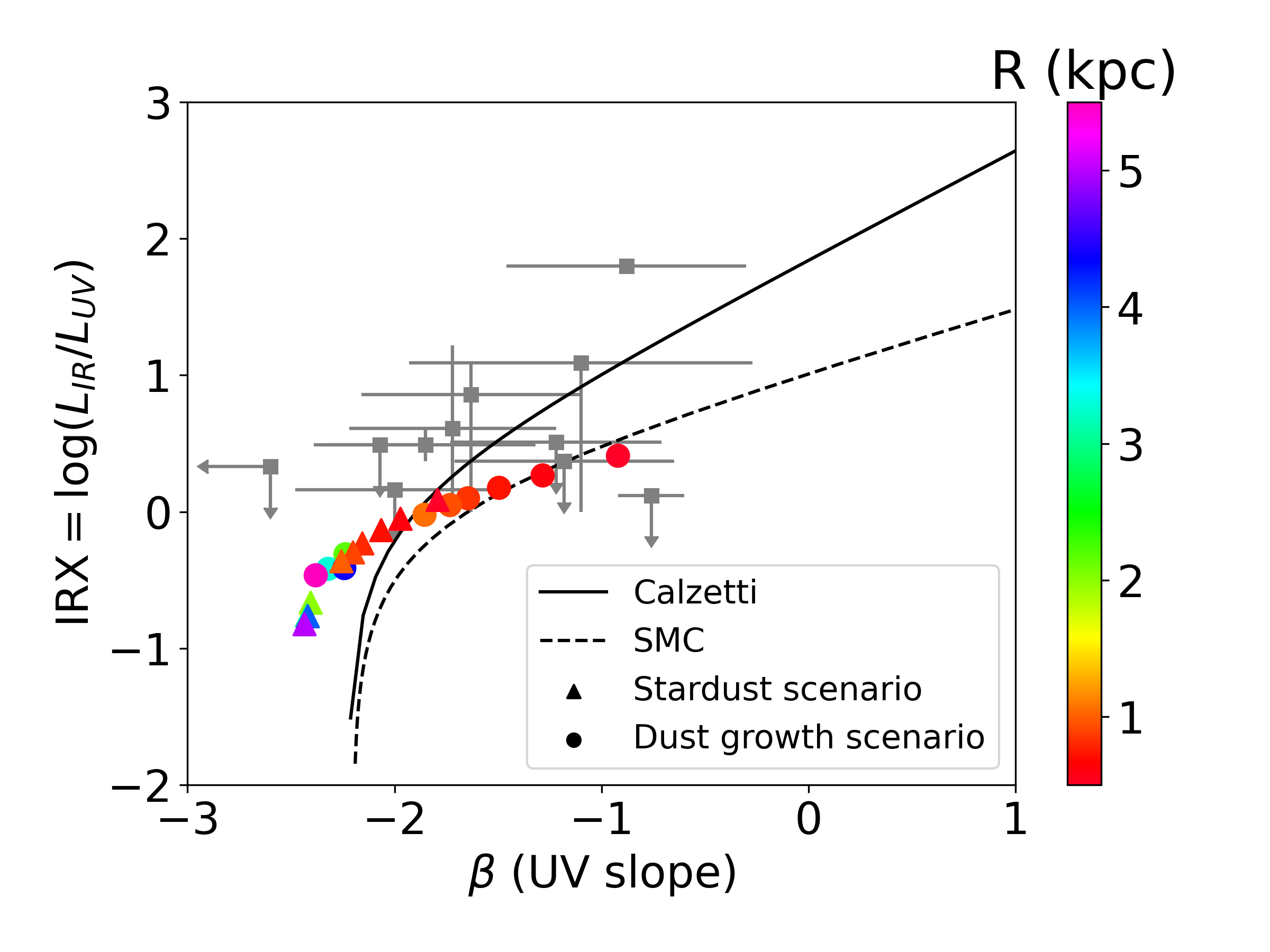}
    \caption{Results for the two-layer geometry.
    Same as Fig.\ \ref{fig:Uni_Att} except for the following points.
    In the top panel, the upper and lower boundaries of the shaded regions correspond to the attenuation curves with $R_{\textrm{young}}=0.5$ and 5 kpc, respectively.
    In the middle and bottom panels, the colour of the points represents the radius of the young sphere ($R_{\textrm{young}}$).}
    \label{fig:TS_Att}
\end{figure}

Next, we present the results for the two-layer geometry.
In the top and middle panels of Fig.\ \ref{fig:TS_Att} we show the effect of changing $R_\mathrm{young}$ on the attenuation curves with the two-layer geometry for the stardust and dust growth scenarios. We find the following behaviours for the attenuation curves:
\begin{enumerate}
    \item When we decrease the radius of the young sphere ($R_\mathrm{young}$), both the stardust and dust growth scenarios show a steepening of attenuation curve. The smaller $R_\mathrm{young}$, the steeper the attenuation curve becomes, because young stars, which dominate the emission at shorter wavelengths, suffer more extinction.
    \item The attenuation curves in the stardust scenario are even steeper than the original extinction curve in the dust growth scenario.
    Therefore, it is difficult to infer the shape of the original extinction curve from the attenuation curve.
\end{enumerate}

In the bottom panel of Figure \ref{fig:TS_Att} we show the IRX--$\beta$ relation for the two-layer geometry. We observe the following features:
\begin{enumerate}
\item As we decrease $R_\mathrm{young}$, both $\beta$ and IRX increase. Since we selectively increase the optical depth for young stars, $\beta$ increases efficiently. This results in a shallow slope on the IRX--$\beta$ diagram, locating some points below the SMC line.
\item The dust growth scenario responds more sensitively to the change of $R_\mathrm{young}$ and has larger IRX and $\beta$ than the stardust scenario. This trend is also seen in the well-mixed geometry.
\end{enumerate}

With the results above, we conclude that selective extinction for young stars steepens the attenuation curves for both scenarios. This is consistent with the results in previous studies \citep[e.g.,][]{Narayanan2018}.


\section{Discussion}\label{sec:discussion}

\subsection{Flattening of attenuation curves}\label{subsec:flattening}
As described in Section \ref{sec:Result}, we observe a flattening of the attenuation curves in the well-mixed geometry for high optical depths (small $R$). This is due to the geometry effect and has already been noted by for example \citet{Witt2000}. We explain our cases as follows.

In the well-mixed geometry with spherical symmetry, the escape fraction of the stellar light (ignoring the effect of scattering) at wavelength $\lambda$ is given by \citep[e.g.][]{Varosi&Dwek1999}
\begin{equation}
    \mathrm{e}^{-\tau_{\lambda,\mathrm{eff}}} \equiv 
    \frac{3}{4\tau_\lambda}\left[ 1-\frac{1}{2\tau_\lambda^2}+\left(\frac{1}{\tau_\lambda}+\frac{1}{2\tau_\lambda^2}\right)\mathrm{e}^{-2\tau_\lambda}\right] ,
    \label{eq:tau_eff}
\end{equation}
where $\tau_\lambda\propto M_\mathrm{d}/R^2$ is the optical depth from the centre to the surface, and $\tau_{\lambda ,\textrm{eff}}$ is defined as the effective optical depth of the system.  
If the optical depth is high ($\tau_\lambda\gg 1$), the above expression is approximated by $\mathrm{e}^{-\tau_{\lambda,\mathrm{eff}}} \sim 3/(4\tau_\lambda )$.
This leads to
\begin{equation}
    \tau_{\lambda,\mathrm{eff}} \sim \ln{\tau_\lambda} - \ln{\frac{3}{4}}\qquad\text{(for high optical depth)}
    \label{eq:tau_eff_ln}
\end{equation}
in the well-mixed geometry.
For the extinction curve in the dust growth scenario, we have $\tau_{0.1~\micron} / \tau_{0.3~\micron}\sim 9$. For high optical depths,
the difference between the effective optical depths at the two wavelengths converges to a constant value as $\tau_{0.1~\micron,\mathrm{eff}}-\tau_{0.3~\micron,\mathrm{eff}} \sim \ln (\tau_{0.1~\micron}/\tau_{0.3~\micron})\sim 2.2$.
Since the attenuation curve is normalized to the value at 0.3~$\micron$, the slope at high optical depth is estimated as $\sim 2.2/\ln\tau_{0.3~\micron}$, which becomes smaller as the optical depth at 0.3~$\micron$ becomes larger.
This explains the flattening of the attenuation curves at high optical depth. In general, if the original extinction curve is steep, the resulting attenuation curve in the well-mixed geometry will be flattened by the above logarithmic dependence.

The saturation of $\beta$ at high optical depth (Section \ref{Res:WellMixed}; Fig.\ \ref{fig:Uni_Att}) in the well-mixed geometry can also be explained analytically as follows. If we assume that the radiation sources have an intrinsic UV slope $\beta_\mathrm{int}$, the change of UV slope after radiative transfer in the well-mixed geometry is described as
\begin{equation}
    \Delta\beta\equiv\beta - \beta_\mathrm{int} = -\frac{(\tau_{\lambda_2,\mathrm{eff}}-\tau_{\lambda_1,\mathrm{eff}})}{\ln{\lambda_2}-\ln{\lambda_1}},
    \label{eq:delat_beta}
\end{equation}
where $\beta$ is the UV slope after attenuation, and $\tau_{\lambda_1,\mathrm{eff}}$ and $\tau_{\lambda_2,\mathrm{eff}}$ are the effective optical depths (equation \ref{eq:tau_eff}) at the wavelengths used to measure $\beta$.
As shown in equation (\ref{eq:tau_eff_ln}), the effective optical depth of the well-mixed geometry is approximately $\tau_{\lambda,\mathrm{eff}} \sim \ln{\tau_\lambda}-\ln{(3/4)}$ for high optical depth ($\tau_{\lambda_1},\,\tau_{\lambda_2}\gg 1$). Thus, equation (\ref{eq:delat_beta}) is reduced to the following expression:
\begin{equation}
    \Delta\beta\simeq
    -\frac{\ln{(\tau_{\lambda_2}/\tau_{\lambda_1})}}{\ln{ (\lambda_2/\lambda_1)}}~~~\text{(for high optical depth)}
    \label{eq:Beta_HOD}
\end{equation}
in the well-mixed geometry.
We observe that, if the optical depth is high, $\beta - \beta_\mathrm{int}$ only depends on the ratio of $\tau_{\lambda_2}$ to $\tau_{\lambda_1}$, which is determined by the extinction curve (not by the attenuation curve). That is, for a given stellar population and an extinction curve, $\beta$ cannot exceed the value described by the above equation. Therefore, as the optical depth increases, $\beta$ is `saturated' in the well-mixed geometry. The steeper the \textit{extinction} curve, the larger values for $\beta$ the UV SED can achieve.

Using equation (\ref{eq:Beta_HOD}), we predict the upper limit of $\Delta\beta \sim 1.5$ in the dust growth scenario, which is roughly consistent with the simulated results for the well-mixed geometry ($\Delta \beta \sim 2.0$). 
For the stardust scenario, though, equation (\ref{eq:Beta_HOD}) predicts $\Delta \beta \sim 0.13$, while the calculation shows $\Delta \beta \sim 1.2$. We interpret this discrepancy as due to the scattering effect we neglect in deriving equation (\ref{eq:Beta_HOD}).

\subsection{Steepening of attenuation curves}\label{subsec:steepening}
We observe some cases of steepening in attenuation curves for both the stardust and dust growth scenarios when the optical depth is low. This is particularly significant in the stardust scenario, which has a very flat extinction curve but shows significant steepening in the attenuation curves. 
There are two main reasons for the steepening of attenuation curves:
\begin{enumerate}
    \item Age effect.
    As mentioned in Section \ref{Res:TwoLayer}, if young stars are more embedded than old stars, far-UV radiation originating from young stars is attenuated more efficiently. As a consequence, the attenuation curve becomes steeper.
    The drastic steepening seen for the stardust scenario in the two-layer model is interpreted as due to this effect. 
    \item Scattering effect.
    Scattering can steepen the attenuation curves in the following two ways. First,
    shorter-wavelength light has a larger chance of being scattered and thus the path length it travels before escaping the galaxy is longer. This also increases the chance of shorter-wavelength light being absorbed  \citep{Goobar2008}. 
    Secondly, the albedo is larger at longer wavelengths (Fig.\ \ref{fig:GrainSizeDis}). This raises the probability that longer-wavelength photons are scattered back into the line of sight. This also steepens the attenuation curve \citep{Witt2000}. 
    As mentioned by \citet{Witt2000}, a shell-like geometry of dust surrounding the stars tends to show the scattering effect prominently. This geometry is similar to our two-layer geometry for the younger component (but note that the age effect is also important for this geometry). 
    The attenuation curves are steeper than the original extinction curves also in the well-mixed geometry, especially at low optical depth. This is interpreted as due to the scattering effect since there is no age effect in this geometry. As the optical depth becomes larger with smaller $R$ in the well-mixed geometry, this effect becomes less important as the flattening effect in Section \ref{subsec:flattening} becomes more prominent.
\end{enumerate}

\subsection{Can the two scenarios be distinguished?}\label{subsec:distinguish}
As shown above, attenuation curves easily become very different from the original extinction curve.
In an optically thin case, scattering can play an important role in steepening the attenuation curves. 
As the optical depth becomes large, the age effect (steepening; Section \ref{subsec:steepening}) or the geometry effect (flattening; Section \ref{subsec:flattening}) becomes more prominent depending on the dust distribution geometry. In particular, the well-mixed geometry could suppress the slope of the attenuation curves by the geometry effect. As a consequence of all the above effects, the attenuation curve slopes cover a very wide range so that the two scenarios (stardust and dust growth) could predict similar attenuation curves despite very different original extinction curves. This suggests that it is not easy to distinguish between the two scenarios (or two different dust sources) based on the slope of attenuation curve.

The two scenarios may cover different regions in the IRX--$\beta$ diagram. In most of the examined cases (especially in the two-layer geometry), the attenuation curves are steeper than the Calzetti and SMC curves. A steeper attenuation curve means that a larger $\beta$ (a redder colour) is more easily achieved. 
This indicates that some data points with low IRX but high $\beta$ are easily explained by our model. Indeed, as observed in Fig.\ \ref{fig:TS_Att}, the dust growth scenario (the steeper extinction curve) more easily explains the observational data points with stringent upper limits for IRX. In other words, the steepness of the original extinction curve has an imprint on the IRX--$\beta$ relation in such a way that the dust growth scenario (the steeper extinction curve case) tends to reproduce the data points that have large $\beta$ (see equation \ref{eq:Beta_HOD} for the mixed geometry) with low IRX.

In fact, equation (\ref{eq:delat_beta}) can be used to explain why similar (but slightly different) attenuation curves in the two scenarios (especially for high optical depth in the mixed geometry; Section \ref{Res:WellMixed}) result in largely different values of UV slope ($\beta$). We rewrite equation (\ref{eq:delat_beta}) as
\begin{align}
        \Delta\beta = -\tau_{\lambda_1,\mathrm{eff}}\frac{(\tau_{\lambda_2,\mathrm{eff}}/\tau_{\lambda_1,\mathrm{eff}})-1}{\ln{\lambda_2}-\ln{\lambda_1}}.
\end{align}
From this equation, we observe that the difference in the attenuation curve slope $\tau_{\lambda_2,\mathrm{eff}}/\tau_{\lambda_1,\mathrm{eff}}$ is amplified by a factor of $\tau_{\lambda_1,\mathrm{eff}}$ in the resulting $\beta$. This explains the reason why we obtain different $\beta$ in the two scenarios at high optical depth even if the attenuation curve slopes are only slightly different.

\subsection{Possible other effects}\label{subsec:clumps}

Although we keep our model simple and minimal to investigate some essential effects on attenuation curves, it is worth mentioning other effects that were not included. We discuss here the effects of clumpiness, stellar age, and dust composition. A recent comprehensive investigation of various effects on the IRX--$\beta$ relation can be found in \citet{Liang2021}.

Clumpiness of the ISM affects the attenuation curve. \citet{Scicluna2015} and \citet{Seon2016} showed that clumpy media predict flattening of the attenuation curve because clumps could reduce the effective optical depth with a given amount of dust, especially at short wavelengths where the optical depth is high \citep{Witt1996,Bianchi2000,Witt2000,Inoue2020}. As shown above, this effect is also caused by the well-mixed geometry; thus, the effect of clumpiness is degenerate to that of mixed geometry. Which of these two effects dominates the flattening of attenuation curve depends on the dust distribution geometry realized in real galaxies. Moreover, previous works such as \citet{Saftly2015} showed that small-scale structures in simulated galaxies can have a relevant impact on the dust energy balance and the attenuation characteristics. Therefore, as mentioned in Section \ref{subsec:galaxy}, we leave the effect of realistic structures, which could be treated by post-processing simulated galaxies \citep[e.g.,][]{Camps2016,Camps2018,Trayford2017,Trcka2020,Liang2018,Ma2019,Vogelsberger2020}, for future work. Note that we realized the desired spatial distribution of dust using discrete particles (Section \ref{subsec:Radiation}). Therefore, it is straightforward to apply our developed scheme to any particle-based simulation data.

The chemical composition of dust also affects the extinction curve \citep[e.g.][]{Draine1984,Pei1992}. Silicate grains can absorb and scatter far-UV light effectively while it is relatively transparent at optical wavelengths, resulting in a steep extinction curve. On the other hand, the extinction curve of graphite is flatter and has a strong bump at 2175~\AA. Flatter extinction curves would result in steeper IRX--$\beta$ relations and could serve to explain some data points with high IRX but low $\beta$. However, the above-mentioned effects that steepen or flatten attenuation curves still largely modify the attenuation curve regardless of the assumed dust composition, covering a large area in the IRX--$\beta$ diagram. This practically overwhelms the detailed assumption on the dust composition.

As mentioned in Section \ref{subsec:galaxy}, the assumed stellar ages are not essential to the conclusion on the attenuation curves as long as the emission at far-UV and that at longer wavelengths are dominated by stellar populations with different ages. 
The stellar ages more significantly influence $\beta$ because of the change in the intrinsic stellar colour.
If we assume, for an extreme case, that all stars have an age of $3\times 10^8$ yr, we find that, under the same dust--stars geometry, $\beta$ increases by $\sim 1$ on average while IRX remains almost the same. Thus, intrinsic red stellar colours could explain the data points lying below the SMC and Calzetti lines in the IRX--$\beta$ diagram. This result is consistent with previous theoretical studies such as \citet{Popping2017},
\citet{Narayanan2018IRXB}, and \citet{Liang2021}. \citet{Burgarella2020}, based on their SED fitting analysis, showed that older LBGs tend to occupy the lower part of the IRX--$\beta$ diagram. 
There could be other interpretations: \citet{Ferrara2017} explained galaxies below the Calzetti and SMC lines on the IRX--$\beta$ diagram by the existence of relatively infrared-dark dust in very dense clouds, of which the emission is overwhelmed by the high-temperature dust near the cloud surface or in the neighbourhood of massive stars \citep[see also][]{Sommovigo2020}.




\section{Conclusion}\label{sec:Conclusino}
We investigate various effects of dust--stars distribution (geometry) on the attenuation curve and the IRX--$\beta$ relation with different dust grain size distributions predicted for galaxies at $z\gtrsim 7$.
Based on \citetalias{Liu2019}, we consider two representative grain size distributions predicted from two cases where the major dust production mechanism is different. These two cases are referred to as the stardust scenario and the dust growth scenario.
In the stardust scenario, the dust abundance is dominated by stellar dust production. The grain sizes are large in this case so that this scenario predicts a flat extinction curve. 
In the dust growth scenario, the dominant fraction of dust mass is formed by the accretion of gas-phase metals in the ISM. In this case, the grain population is dominated by small grains, so that a steep extinction curve is predicted. The main purpose of this study is to investigate if these two different scenarios produce different attenuation curves or not. We also investigate the IRX--$\beta$ relation, which also reflects the attenuation properties.

To calculate attenuation curves, we consider two different dust geometries that include important mechanisms of altering the attenuation curve slope: the well-mixed geometry and the two-layer geometry. In the well-mixed geometry, we distribute stars and dust uniformly in a sphere with a galactic radius of $R$. In this model, we adjust the optical depth of the system by changing $R$.
In the two-layer geometry, we locate all young stars in the centre of the sphere, while we distribute old stars uniformly in the sphere of radius $R_\mathrm{old}$. We separate the dust into the young sphere and old sphere, respectively, with radii $R_{\textrm{young}} \leq R_{\textrm{old}} $. We change $R_{\textrm{young}}$ to adjust the optical depth for young stars (see Fig.\ \ref{fig:Model}).

For the well-mixed geometry, the attenuation curves are steep at low dust optical depth in both dust scenarios because of the scattering effect. The slope of attenuation curve tends to decrease as the optical depth increases, reaching a similar slope to the SMC or Calzetti curves. This flattening occurs because a certain fraction of short-wavelength light always escapes by the geometry effect.
Increasing the optical depth raises both IRX and $\beta$.
The steep attenuation curves induce a significant increase of $\beta$ for $R\lesssim 4$ kpc. This increase saturates at $R \lesssim 1$ kpc because of the above-mentioned escaping effect (or the geometry effect). For a fixed galaxy radius, the dust growth scenario has larger $\beta$ because of the steep extinction curve.

In the two-layer geometry, we observe steep attenuation curves. This is because of the age effect; that is, shorter-wavelength light emitted by young stars suffer more attenuation. The attenuation curve becomes steeper as we increase the optical depth for young stars. In the IRX--$\beta$ diagram, the dust growth scenario tends to show larger IRX and $\beta$ under the same $R_\mathrm{young}$ than the stardust scenario. The steep attenuation curves also explain data points that have large $\beta$ but relatively suppressed IRX.

The different effects described above
simultaneously contribute to the variation of attenuation curves. The resulting attenuation curves show a wide range of slope. Thus, it is difficult to discriminate between the two dust scenarios (i.e.\ the major dust sources) only from the attenuation curve. On the other hand, the two dust scenarios are more separated in the IRX--$\beta$ diagram. The dust growth scenario tends to predict larger $\beta$ for a fixed value of IRX than the stardust scenario.

We conclude that the two possible dust origins (stellar dust production and dust growth) can lead to similar attenuation curves, even though they are characterized by strongly different extinction curves. We need dust emission data, such as the IRX--$\beta$ relation, in order to distinguish between the two dust origins. Alternatively, we need to measure extinction curves, not attenuation curves, using bright background sources as a key to discriminate which of the two dust sources dominates the dust production at high redshift.

\section*{Acknowledgements}

We thank Yu-Hsiu Huang, Yun-Hsin Hsu, and the anonymous referee for useful discussions and comments.
HH thanks the Ministry of Science and Technology for support through grant
MOST 107-2923-M-001-003-MY3 (RFBR 18-52-52006) and MOST 108-2112-M-001-007-MY3,
and the Academia Sinica
for Investigator Award AS-IA-109-M02.

\section*{Data availability}

The data underlying this article will be made available in Figshare at \url{https://doi.org/10.6084/m9.figshare.13059845.v1}.




\bibliographystyle{mnras}
\bibliography{reference}






\bsp	
\label{lastpage}
\end{document}